# Electron Density Measurement with Hairpin Resonator in Low Pressure Plasma

Xingchen Fan, Patrick Pribly, Troy Carter
University of California, Los Angeles
Plasma Science and Technology Institute
e-mail: xcfan@ucla.edu

**Abstract**

Hairpin probes measure electron density in plasmas where other techniques are unsuitable. Previous hairpins were constructed as a quarter-wave resonant structure typically made from a folded piece of wire. The present work conducts the hairpin measurement in high electron density, up to approximately $10^{12}$ / cm$^3$, corresponding to a plasma frequency of about 9 GHz. A series of four probes is described, together with an easily reproducible implementation of the associated microwave electronics using commercial off the shelf components that are inexpensive compared to the network analyzer that is typically required. Measuring higher densities requires increasing the resonant frequency, but we were unable to accurately fabricate the required quarter wavelength structure on the scale of 4 mm. Toward this end we explore operating a larger quarter-wavelength structure at its 3d harmonic (3/4 resonator). Correction coefficients are described for both the plasma sheath effect and the wire coating. Measurements are taken in Argon at pressures below 20 mTorr. Results are compared with Langmuir probe measurements.

**1. Introduction**

The hairpin resonator has been used to measure plasma density for over 40 years, and was first introduced by Stenzel in 1976 [1]. For years, incremental developments have improved the precision, enhanced the spatial resolution, and extended the application to a wide range of plasmas. Use of a hairpin probe, named after its physical characteristics, is practical in many kinds of plasma, such as the pulsed ICP discharge [2], the nitrogen afterglow [3], and the dust particles [4]. The hairpin is sensitive to changes of multiple parameters of plasma. For example, due to its behavior at higher pressures, it can be used to measure the collision frequency in moderate pressure plasma (1 - 10Torr) [5]. When it measures the electron density in the higher pressure (>1 Torr), a collisional correction is put forward [6]. Immersed in a magnetized plasma, the probe has two resonances occurring around its resonant frequency in vacuum due to the plasma frequency and the cyclotron resonance and their interaction[7].

The apparatus setup is critical. Researchers describe directly coupled and indirectly coupled resonators; in the first, coaxial cables with the microwave signal electrically connect directly to the hairpin, while in the second inductive couplers are located near the hairpin resonator[8]. In addition, measurements made in transmission or reflection modes have been described [9]. Transmission mode uses two signal lines, with the microwaves transmitted across the hairpin from an input to an output line, while reflection mode uses a single line with both the launched signal and its reflection propagating down the same coax. In much of the literature researchers prefer reflection mode, possibly because it only needs the one signal line. We tested the four possible combinations, and for the high frequencies we describe, determined that indirectly coupled transmission mode configuration gave the best performance in terms of well-defined resonance at the highest frequency. For any hairpin probe the resonant frequency is not unique. In principle, any wave whose node and antinode stay at two ends of the hairpin is a standing wave, therefore a resonance occurs. We select the waves whose quarter and 3-quarter wavelengths equal to the total length of the hairpin as the standing waves that evoke the needed resonance.

In the following sections, the hairpin probe theory is introduced briefly. Based on the theory, a diagnostic system and hairpin probes with different dimensions are built to measure the electron density in the low pressure (20mTorr). The hairpin probes are tested in an inductively coupled plasma (ICP) with Argon. In the presence of the sheath and the coating, different measurements require corresponding corrections. In the end, the corrected measurements are compared with other plasma diagnostic techniques.

## 2. Hairpin Theory

The hairpin is a "U" shaped antennae and behaves as a transmission line. More specifically, it is a quarter-wavelength transmission line with one end shorted and the other open. An adjustable microwave frequency is coupled to the hairpin near the shorted end using a small loop formed at the end of a coax. In transmission mode a receiving loop of the same geometry is position on the other side of the loop. When the signal frequency is equal to a multiple of the resonant frequency of the hairpin, a standing wave occurs on the hairpin. Experimentally we find that in this condition the receiver has the largest signal. The standing wave is detected either in the transmission mode or reflection mode, which will be discussed in the next section.

The resonant frequency in a media is given approximately by the formula

$$f_r = \frac{c}{2(2l+w)\sqrt{\epsilon}} \quad (1)$$

where $c$ is the speed of light, $l$ is the length of one leg of the hairpin, $w$ is the width between the two legs, and $\epsilon$ is the dielectric constant of the surrounding media.

When the hairpin is in vacuum, the resonant frequency $f_0$ is

$$f_0 = \frac{c}{2(2l+w)} \quad (2)$$

Without any external magnetic field, the real component of the dielectric constant of the plasma is

$$\varepsilon = 1 - \frac{f_p^2}{f^2} \quad (3)$$

where $\omega_p = 2\pi f_p = \left(\frac{4\pi n e^2}{m}\right)^{1/2}$ is the plasma frequency. $e$ and $m$ are the charge and mass of an electron respectively, and $n$ is the electron density.

Operating at frequencies $f > f_p$, equation (3) is always valid. Plug equation (3) into equation (1) to derive the relation between the frequencies under the condition that $f = f_r$

$$f_r^2 = f_p^2 + f_0^2 \quad (4)$$

where $f_0$ is the resonant frequency of the hairpin in vacuum.

A sheath will form around the legs of the hairpin when placed in the plasma. The sheath is a cylindrical volume centered on the axis of each wire and to a first approximation free of electrons which, therefore, leads to an underestimation of electron density between two legs of the hairpin. To accurately measure the density, a correction is introduced as discussed in Sands et al [6].

Then, equation (4) becomes

$$f_r^2 = \xi_s f_p'^2 + f_0^2 \quad (5)$$

where $f'_p$ is the corrected plasma frequency and $\xi_s$ is the correction factor. The sheath is regarded as a capacitance in series with the resonator, therefore the corrected frequency has a new relation with the hairpin dimension from which the correction factor is eventually derived. $\xi_s$ given by the formula [6]

$$\xi_s = 1 - \frac{f_0^2}{f_r^2} \frac{\left[\ln\left(\frac{w-a}{w-b}\right) + \ln\left(\frac{b}{a}\right)\right]}{\ln\left(\frac{w-a}{a}\right)} \quad (6)$$

where $a$ is the radius of the hairpin wire and $b$ is the radius of the cylindrical sheath around the wire. Here we assume the sheath extends one electron Debye length, $\lambda_D = (\epsilon_0 T_e / n_e e)^{1/2}$, out from the surface of the hairpin wire. The wire we use to construct the hairpin is of radius $a = 0.25$mm, and in these plasmas the Debye length is about 13 to 40 microns.

We assume these low-pressure plasmas the collision correction is neglectable. The formula for the corrected electron density is then

$$n_e = \frac{\pi m_e}{e^2} \frac{f_r^2 - f_0^2}{\xi_s} \qquad (7)$$

## 3. Experimental Setup
*3.1 Microwave Circuit*

A microwave circuit has been designed and constructed to realize the hairpin measurement. The system consists of a microwave oscillator, a coupler, a mixer, and an amplifier. A schematic is shown in Fig. 3. The oscillator is an HMC588LC4B from Analog Devices with a frequency range from 8.0 to 12.5 GHz and output power of 5 dBm. We use an FPC06074 directional coupler from Knowles Dielectric Labs which has a coupling coefficient from 10 to 12 dB for the frequency range 8 to 12 GHz. The pivotal part of the system is the mixer which we choose to use SIM-153LH+ from Mini-Circuits. It is a Level 10 (LO Power +10 dBm) mixer for 3.2 to 15 GHz signals. In order to make the LO port receive enough power, we use the amplifier CMD157P3 from Custom MMIC to give a +24 dB boost to the signal between 6 and 18 GHz. Thanks to X-Microwave [11], all the parts are made into modules and ready to assemble. The whole system fits into a 1.34*2.83 inches machined housing also available from the same vendor.

A 5V DC source powers the oscillator and the control voltage ranges from 0 to 13V. A 0-10V ramp signal, sweeping in 10 ms, serves as the control voltage, tuning the approachable output signal from 7.5 to 12.9 GHz. The curve of frequency vs. tuning voltage is provided in Fig. 1. As diagramed in Fig.3, the signal from the oscillator directly goes into the coupler, and a -7 dBm coupling signal is sent into the RF port of the mixer while the major part of the signal is sent into the input loop that couples the signal onto the hairpin. Any signals picked by the pick-up loop from the hairpin are sent into the LO port of the mixer through the amplifier. Fig. 3 and 4 are the schematic diagram and the real picture of the system.

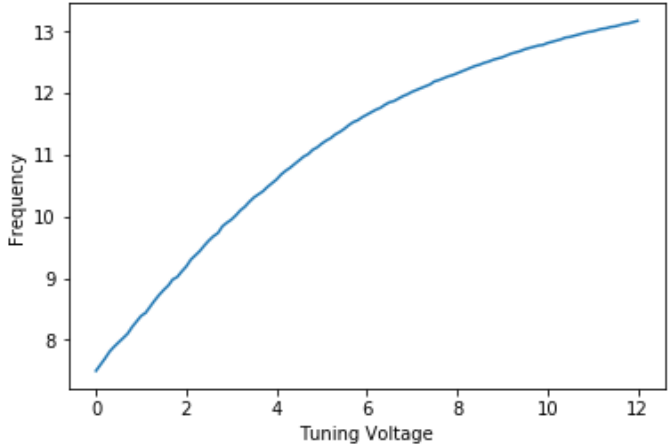

(Fig. 1 Output frequency - tuning voltage diagram of the oscillator )

Directly coupled and indirectly coupled resonators are two types of probe that we can choose from with this system. The directly coupled resonator requires the loop to be soldered onto the resonator, while the indirectly coupled resonator floats without the contact with the loop. Although the directly coupled resonator attained a stronger signal, a sheath will usually form around the now-grounded probe which would greatly interfere with the measurement, particularly in the case of plasmas generated by RF [8]. We use the indirectly coupled resonator, meaning the hairpin doesn't contact the feed coax, so that the hairpin is totally floating in the plasma. In low electron density plasma, researchers tend to use the reflection mode because it only requires one loop which is easier to construct and slightly smaller. However, the transmission mode has a higher density measurement limit. Specifically, by the measurement of Piejak et al [9], the measurable density is as high as $1.3 \times 10^{12}$ /cm$^3$. Moreover, according to our observation, when using the reflection mode, the resonant frequency of the hairpin is almost lost in the standing wave signal arising from impedance mismatch at the coupling to the hairpin. This effect is worse when the frequency is high and the transmission line is long. Transmission mode on the other hand only relies on accurate termination at the receiving end, which is easy to accomplish.

The semi-rigid copper coaxial cables UT-034M from Amawave [12] with OD = 0.034 inches(0.87 mm) are employed to send and receive the microwave signals. Attenuation in these cables is about 5 dB/m, among the lowest of this type of cable. The input and pick-up loops are at the end of the coaxial cables and made by soldering the inner conductor of the coax to the shield. Before fixing the hairpin between the two loops, the loops are adjusted to be in the alignment to give the best transmission and to minimize the resonance of the loops themselves. Finally, we use epoxy [13] to keep them in position (Fig. 2).

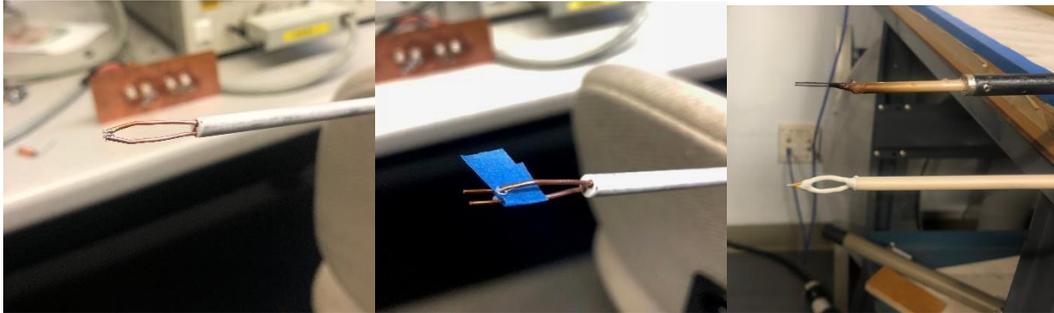

Fig. 2 Hairpin construction. First panel: the transmit and receive loops are carefully aligned. Second panel: the hairpin, consisting of a bent piece of wire with legs as straight as possible is positioned and tested for resonance. Third panel: two completed hairpins are shown. The top one has been used in a high power high density plasma.

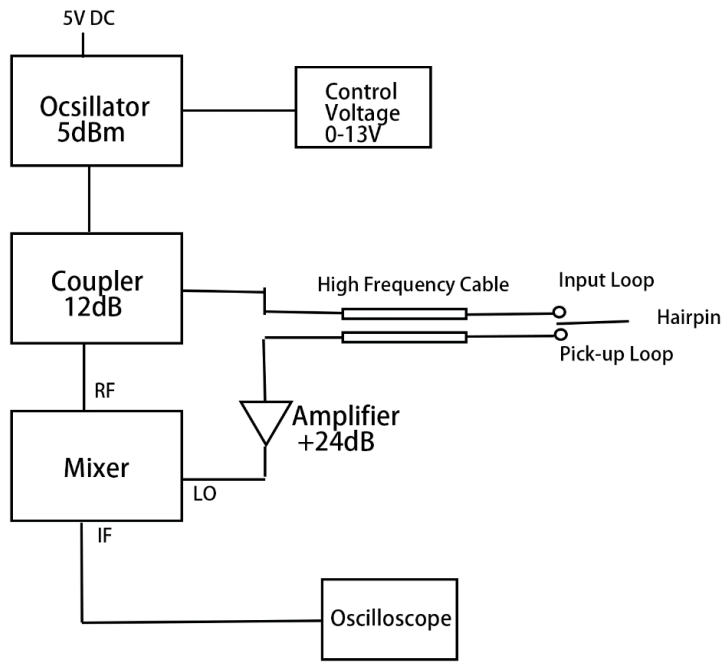

Fig. 3 Schematic diagram of the system. Components are described in the text, and most of the active microwave parts are available as commercially pre-assembled building blocks as part of a system from X-Microwave [11], see Fig. 4..

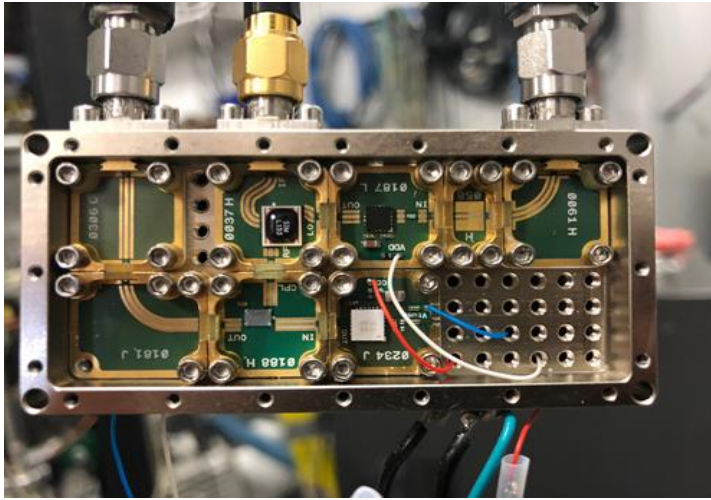

Fig. 4 Microwave electronics as implemented using X-Microwave components. The individual commercially available modules interconnect using specially designed impedance-matched bridges. The left-most SMA connector is the transmitter**,** the center connector is the mixer output, and the right one is the receiver. The high frequency cables connecting the box to the probe have less than 1.4 dB attenuation.

*3.2 Hairpin Design and Pre-measurement*

As discussed above, the hairpin we are using has no direct contact with any loops or coaxial cables. Its potential is isolated from the ground and the signals we send in, thus the sheath correction is simplified.

The quality of the hairpin depends on the material and the dimensions. In principal, the material with higher conductivity would constitute a hairpin with less loss and more sensitivity. We have tested copper and silver as the materials because their conductivities are sufficiently high and they are stable in the plasma environment that we are probing. Although the resonant frequency is determined by the total length of the hairpin, the Q factor is affected by the ratio of l and w (leg length and width between two legs) [6]. We verified this relation by comparing three hairpins with almost the same resonant frequency in vacuum but different *l* to *w* ratios. Results from the network analyzer are shown in Fig. 5, and using these we can calculate the Q factors for three hairpins. Their dimensions and the Q factors are in Table 1.

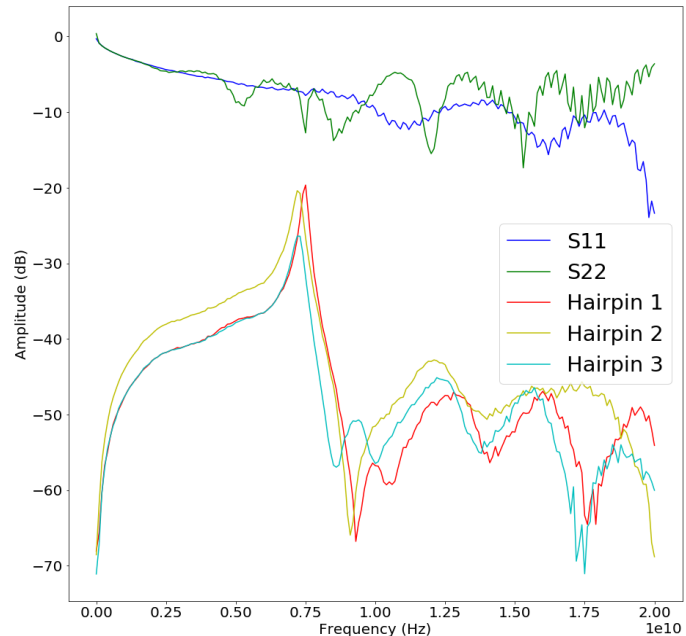

Fig. 5 Comparison of hairpins with dimensions given in Table 1.

| Dimensions | Hairpin 1 | Hairpin 2 | Hairpin 3 |
|---|---|---|---|
| Width ($w$)/mm | 1.78 | 2.90 | 3.90 |
| Length ($l$)/mm | 9.25 | 9.15 | 8.30 |
| Ratio ($l/w$) | 5.20 | 3.16 | 2.13 |
| $f_0$/GHz | 7.5 | 7.2 | 7.2 |
| $Q$ Factor | 37.5 | 24.0 | 14.4 |

Table 1. Dimensions and quality factor $Q$ of similar frequency hairpins for the in Fig. 5

    The higher aspect ratios give larger $Q$ values, which helps both in detectable signal and density resolution. Ideally, in order to build the most sensitive hairpin, the ratio should be as high as possible. Therefore, decreasing the width and increasing the length can optimize the ratio. However, the width is limited by both the physical characteristics of the material and the need to avoid overlapping sheaths for the legs. On the other hand, as the length increases, the resonant frequency drops and the perturbation caused by the hairpin becomes larger. In practice, using this method we were unable to construct hairpins with adequate $Q$ for quarter-wave resonant frequencies higher than about 11 GHz, for a highest measurable density of about $10^{12}$/cm$^3$.

    A top measurable density of $10^{12}$/cm$^3$ is insufficient for the higher density plasmas typically produced at the BaPSF facility. With the goal of measuring densities up to $10^{13}$/cm2, we also explored the possible usage of the three-quarter-wavelength resonant frequency. Thus, a hairpin we could construct with a lowest resonance at 11 GHz could potentially be over-moded to operate at a much higher frequency. Rather than buying another set of higher frequency microwave hardware, we made measurements for a larger hairpin such that our available range of frequencies represented the 3d harmonic. The hairpin for this exploration is 22.83 mm long and 1.3 mm wide. The pre-measurement by the network analyzer shows the behavior of this frequency in Fig. 6. Its advantage is that, for a given hairpin, the frequency can be three times higher, and given a resonant frequency, the $Q$ factor is larger than the ¼ resonator shown for comparison; this is consistent with the higher $l$ to $w$ aspect ratio. However, as is apparent from the red trace in Fig. 6, the transmission response is somewhat lower than that at the quarter-wavelength frequency.

    No matter what dimensions of the hairpins we choose, our $Q$ values are below 100 when the resonant frequencies are higher than 8.0GHz. The hairpins reported by Sands et al [6] and Peterson et al [5] have $Q > 100$, but their resonant frequencies fall in the range of 2.0 - 3.1GHz. The compromise of accuracy is unavoidable when pursuing higher measurable frequencies.

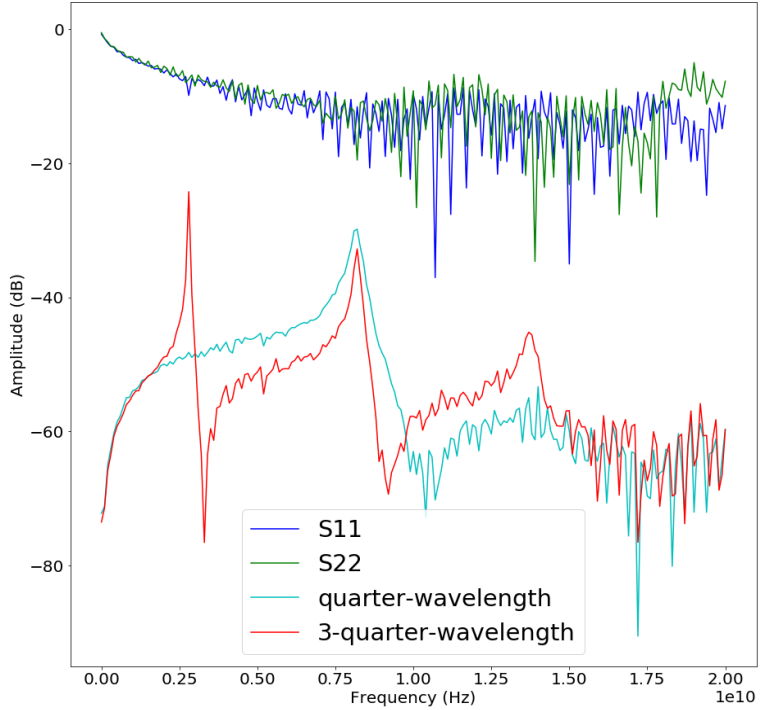

Fig. 6. Comparison of the quarter-wavelength and 3-quarter-wavelength hairpins. The two resonators are constructed to overlap at $f_0$ = 8.2 GHz. The lower two traces are network analyzer measurements in transmission mode (S22). The quarter-wavelength hairpin has Q ≈ 20 at $f_0$, while 3-quarter-wavelength version has Q ≈ 41. Note that the top two traces are measurements of S11 and illustrate the difficulty with using a reflection measurement scheme at these frequencies.

*3.3 Correction for Presence of a Coating on the wire*

      The copper wire we use has a thin insulation layer on the surface. This coating while inside the plasma serves as a capacitor whose effect we may have to consider. Xu et al introduced a correction on the effect by a dielectric protection film in high pressure [10]. Xu used the relation between the width of reflection spectrum and plasma characteristics to obtain the density on which the coating correction he developed is valid, but we are in the low-pressure plasma and use different relation to derive the electron density. As a result, our coating correction should be different from him.

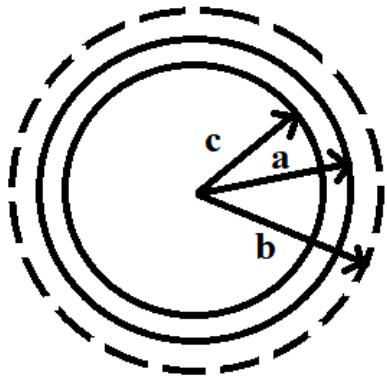

Fig. 7 Cross section of the copper wire. Here $c$ = the wire radius, $a$ = radius with insulation layer, and $b$ = radius of plasma sheath.

Shown in Fig. 7, *a* and *b* are insulation and sheath radii, and *c* is the wire radius without the coating. For a hairpin in a uniform plasma, the effective capacitance per length is

$$C_{eff} = \frac{\varepsilon}{4\ln\left(\frac{w-c}{c}\right)} \quad (8)$$

where $\varepsilon$ is the effective permittivity, following equation (1). The ratio of vacuum and plasma resonance frequencies, $f_0$ and $f_r$, give the dielectric value according to $\varepsilon = (f_0/f_r)^2$. Rewrite equation (7) as

$$C_{eff} = \frac{(f_0/f_r)2}{4\ln\left(\frac{w-c}{c}\right)} \quad (9)$$

The hairpin, the cylindrical sheath, and the cylindrical coating are capacitors in series. Similar to the derivation by Sands et al [7], the corrected capacitance per length of the hairpin is

$$C_{corr} = \frac{\varepsilon_p}{4\left[\ln\left(\frac{w-b}{b}\right)+ \varepsilon_p\left(\ln\left(\frac{w-a}{w-b}\right)+\ln\left(\frac{b}{a}\right)\right)+\frac{\varepsilon_p}{\varepsilon_c\left(\ln\left(\frac{w-c}{w-a}\right)+\ln\left(\frac{a}{c}\right)\right)}\right]} \quad (10)$$

where $\varepsilon_p$ is the actual permittivity of the plasma that complies to the dispersion relation, equation (2), and $\varepsilon_c$ is the permittivity of the coating material. The coating thickness is about 30 um and the material is polyvinyl formal (Formvar) whose $\varepsilon_c$ = 2.1. Equating equation (8) and (9), and substituting $\varepsilon_p$ with (1 - $(f_p'/f_r)^2$), the corrected plasma frequency is given by

$$f_p'^2 = \frac{f_r^2 - f_0^2 \frac{A+\ln\left(\frac{w-b}{b}\right)}{\ln\left(\frac{w-c}{c}\right)}}{1-(f_0/f_r)2\frac{A}{\ln\left(\frac{w-c}{c}\right)}} \quad (11)$$

where A is

$$A = \left[\ln\left(\frac{w-a}{w-b}\right) + \ln\left(\frac{b}{a}\right)\right] + \frac{1}{\varepsilon_c}\left[\ln\left(\frac{w-c}{w-a}\right) + \ln\left(\frac{a}{c}\right)\right] \quad (12)$$

Hence, the corrected electron density is

$$n_e = \frac{\pi m_e}{e^2} f_p'^2 \quad (13)$$

## 4. Experimental Results and Discussion

*4.1 Experimental Results*

In the inductively coupled plasma (ICP), higher RF power coupled into the plasma produces higher the electron density in an almost linear relation. For this work, we used a mixer with a single output (an "I/Q" mixer would be preferable but was not available). Consequently output signals from the IF port of the mixer have phase information encoded in them as the frequency is swept, as seen by the oscillating waveforms shown in Fig. 8. This figure is a composite of the waveforms taken at different RF powers. To determine the resonant frequency, we select all the local maximums of the absolute value of the waveform, and draw the profiles of these over the frequency range. Then, we fit an offset Gaussian function to this envelope and locate the center of the Gaussian as the resonant frequency. Fig. 9 shows such a fit for an RF power of 230W.

Four hairpins with different dimensions and materials are tested in an inductively coupled plasma (ICP) chamber. Information about these hairpins is in Table 2. The tests were done at different times over a period of a month. Measurements on the four hairpins at different plasma power are shown in Fig. 5 to Fig. 13. Resonant frequencies are pinpointed by Gaussian fitting. All the data are acquired at the center of the chamber. Pressure in the chamber is 20mTorr, which is sufficiently low that corrections due to collisions are not necessary.

| Probe | Length (l) /mm | Width (w) /mm | Vacuum resonant frequency ($f_0$) /GHz |
|---|---|---|---|
| quarter-wavelength Copper Hairpin 1 | 7.96 | 1.30 | 9.06 |
| quarter-wavelength Copper Hairpin 2 | 7.00 | 1.35 | 9.73 |
| quarter-wavelength Silver Hairpin | 6.90 | 0.92 | 10.33 |
| three-quarter-wavelength Hairpin | 22.83 | 1.30 | 9.19 |

Table 2 Dimensions and resonant frequencies of four hairpins tested in the ICP chamber. Probes #1-3 are ¼ resonant structures, while probe #4 operated at ¾ resonance.

Signal loss is always an issue for all the hairpin probes. In our case, we find a fairly abrupt signal loss above a resonant frequency of 11.5 GHz. An extra amplifier needs adding to enhance the signals due to the resonance above the system noise. Unfortunately, some of the noise is also amplified, making data analysis more difficult. For the analysis of these cases we arbitrarily set the signal outside of the range where the resonance occurs to zero. In the case of the 3-quarter-wavelength copper hairpin, all the signals reduce to below the noise level once the frequency surpasses 11.5GHz. The resonance is no longer distinguishable even with the auxiliary amplifier.

Simply comparing the frequency spectra of the four hairpins (Fig. 10), the *Q* value of the 3-quarter-wavelength hairpin is higher than those of both quarter-wavelength hairpins, as expected. However, the resonant peak for the three-quarter-wavelength hairpin drops below the noise when the plasma power goes higher than 600W. This is already a surprise that this hairpin works as a density probe in the plasma with such a power. As the power goes up, the signals received dominate by excited waves that match the lower resonant frequency of the hairpin. Between the two quarter-wavelength hairpins, the one with a lower vacuum resonant frequency has a higher *Q* value. Then, comparing Fig. 8 and Fig. 12, although the silver hairpin has a higher vacuum resonant frequency, the *Q* factor is higher than that of its copper counterpart because of its bigger *l* to *w* aspect ratio and silver's higher conductivity. However, the higher vacuum resonant frequency corresponds to severe signal loss at lower density.

Electron densities shown in Fig. 11 are calculated by applying equation (6), without correction, to the resonant frequencies measured by the microwave system. The probes essentially agree with each other, to within the measured power. Two curves of the quarter-wavelength copper hairpin match with each other with a very slight discrepancy. And both the three-quarter-wavelength copper hairpin and quarter-wavelength silver are lower than the quarter-wavelength copper hairpin. The discrepancies grow somewhat as the plasma power increases. The 3-quarter-wavelength hairpin is almost two times longer than the quarter-wavelength one, so part of its perturbation to the plasma may be the reason for its lower inferred density, although more work is needed to pin this down.

For an electron density of $10^{11}$/cm$^3$ and temperature 3 eV, the Debye length is approximately 0.004 cm. This value is inversely proportional to the square root of the density. According to the discussion of Sands et al [6], the sheath correction factor becomes trivial when probe separation *w* is much greater than *b*, the radius corresponding to conductor radius + insulation + Debye layer thickness. The sheath effect would considerably affect the result because neither the wire+insulation radius *a* nor the width *w* is much larger than the thickness of the sheath ($\lambda_D = b - a$).

There is no coating on the silver hairpin, so we only apply the sheath correction to its measurements. Assuming $b = a + \lambda_D$, solve equation (5) iteratively for the sheath correction factor $\xi_s$. For each resonant frequency, $\xi_s$ has a corresponding value. As the resonant frequency increases, indicating the density increase, $\xi_s$ approaches unity, or no correction. Accordingly the sheath effect correction makes less difference when the density is higher, which agrees with the observation of Piejak et al.[9] in a lower electron density range, 4 - $30 \times 10^{10}$/cm$^3$. After solving $\xi_s$ for all the frequencies, we again apply equation (6) to obtain the corrected electron density. As for the copper hairpins, both the coating and the sheath corrections should be involved. Assuming $\lambda_D = b - a$, solve equation (12) and (13) iteratively for the corrected electron density.

      Fig. 12 presents the corrected density and raw density vs. plasma power for the quarter-wavelength hairpin 1. The difference between the corrected and uncorrected densities varies from 9% to 48%. As the density increases, the difference becomes smaller in percentage. Since all the three copper hairpin probes have similar width *w*, the correction gives almost the same percentage increase on the density. However, the sheath correction makes a 6% to 29% difference to the measurements of the silver hairpin probe due to its lack of coating. Fig. 13 shows the sheath corrected density measured by four hairpin probes vs. plasma power. Langmuir probe measurements based on ion saturation current are also shown for comparison. The correction makes the discrepancies among the coated copper hairpin probes smaller than 15% when plasma power is lower than 600w, while the difference between silver hairpin probe and the one-quarter-wavelength hairpin 1 curves remains. The silver hairpin measurements are about 16 to 46% lower than the hairpin 1. Comparing the hairpin 1 measurements and the Langmuir probe measurements, the latter are about 61 to 48% less than the former in a plasma power range of 200 - 900w.

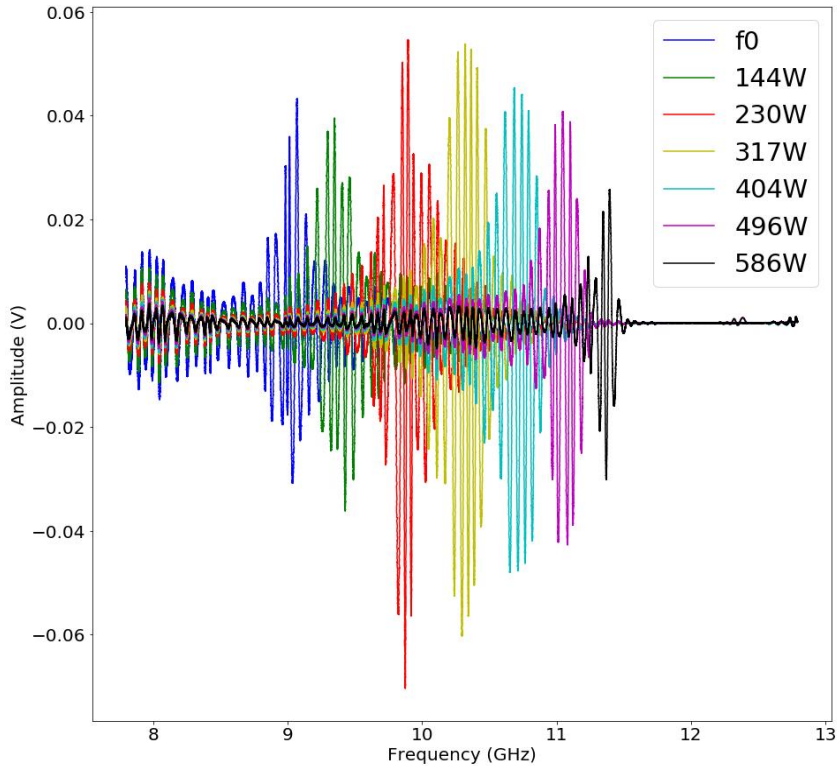

Fig. 8 Mixer output signals for different plasma densities as controlled by varying the plasma power. Power is indicated in the legend. The horizontal axis is the microwave frequency, swept from about 8 to 13 GHz.

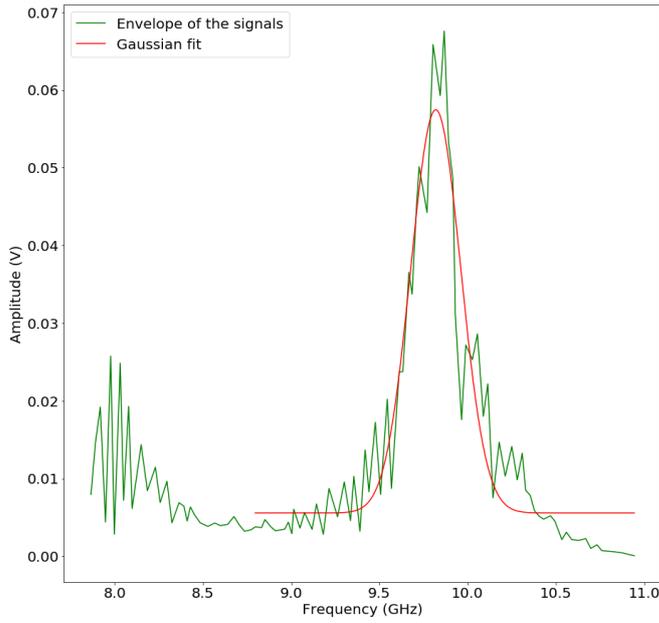

Fig. 9 Envelope and Gaussian fitting of the signals of the quarter-wavelength copper hairpin #1 at plasma power 230W. An offset Gaussian is fitted to the absolute value of the mixer output for this probe shown as one of the traces in Fig. 8.

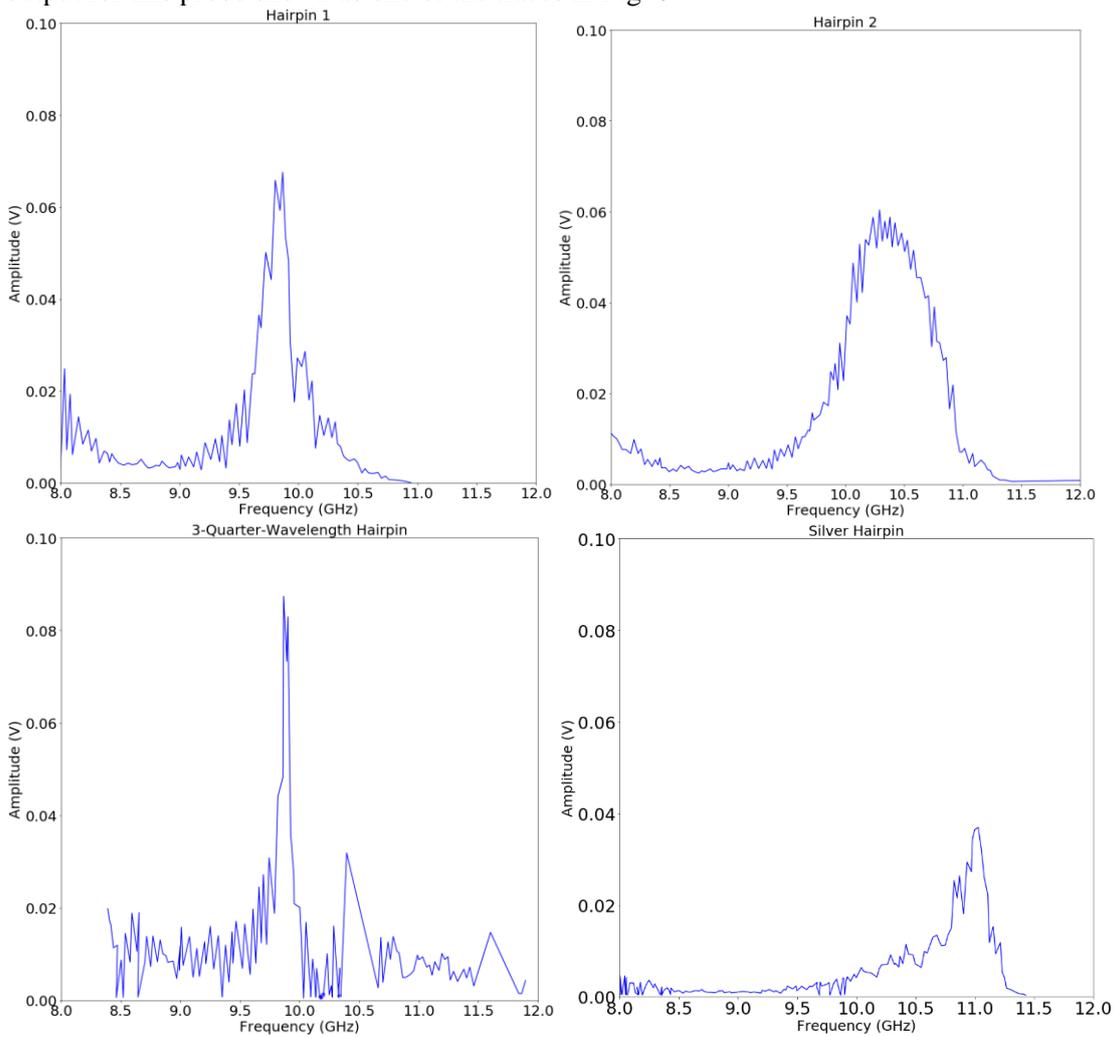

Fig. 10 Frequency spectra of four hairpins in 230W plasma. Panels *a-c* show results from ¼ resonance probes described in Table 2, and panel *d* from the ¾ resonance probe.

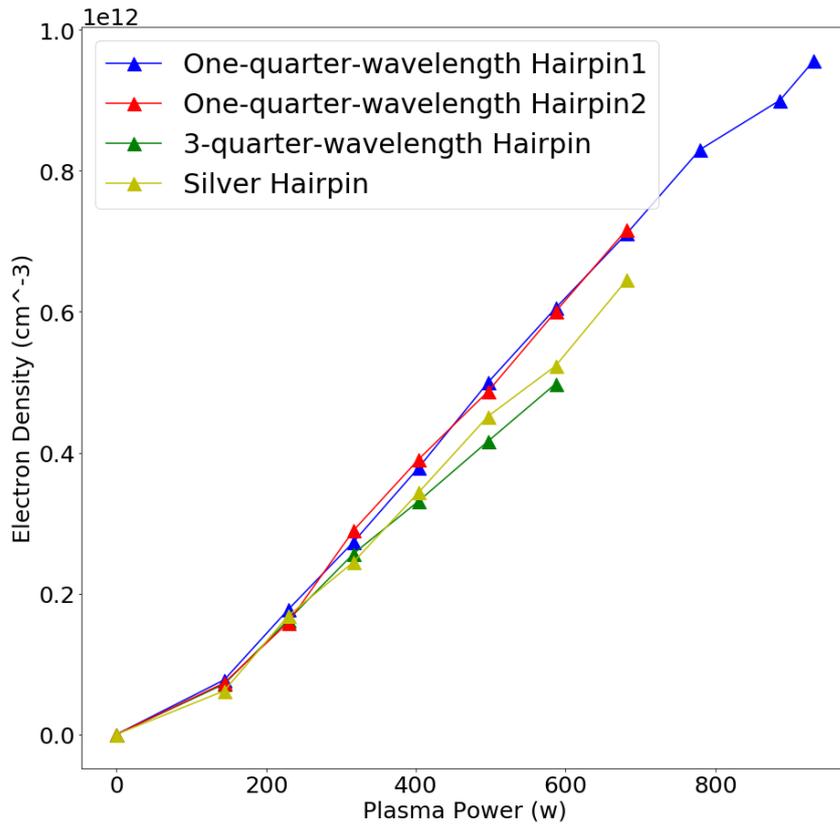

Fig. 11 Raw Density vs. plasma power. Results are shown for the four probes described in Table 2. Densities are converted from the resonant frequencies measured on the hairpin probes without sheath effect correction.

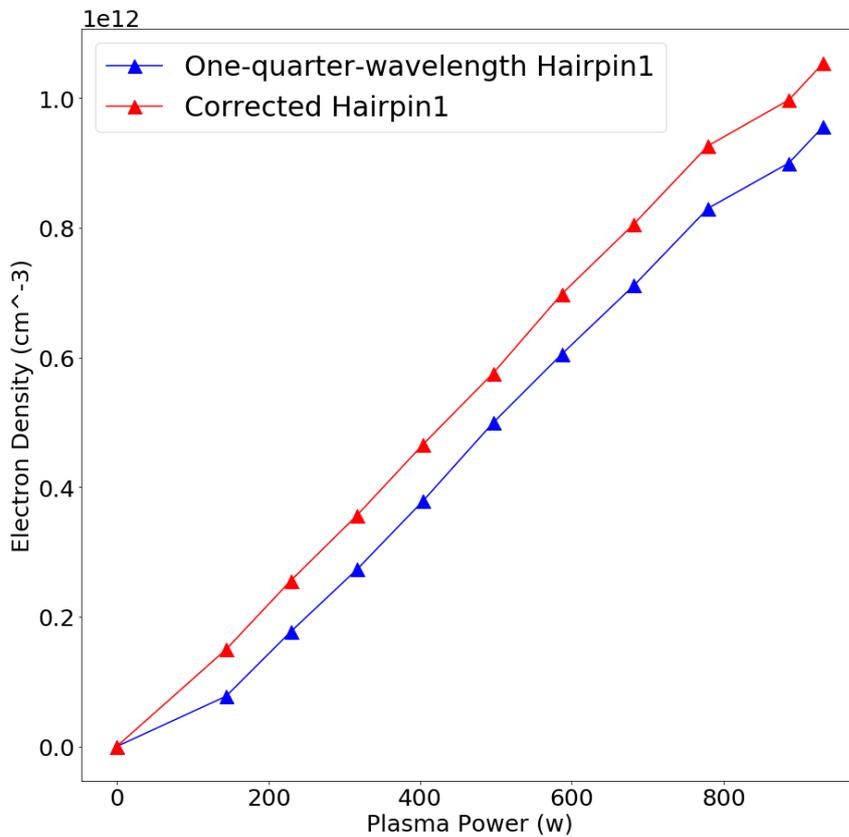

Fig. 12 Raw density and corrected density of the quarter-wavelength copper hairpin 1.

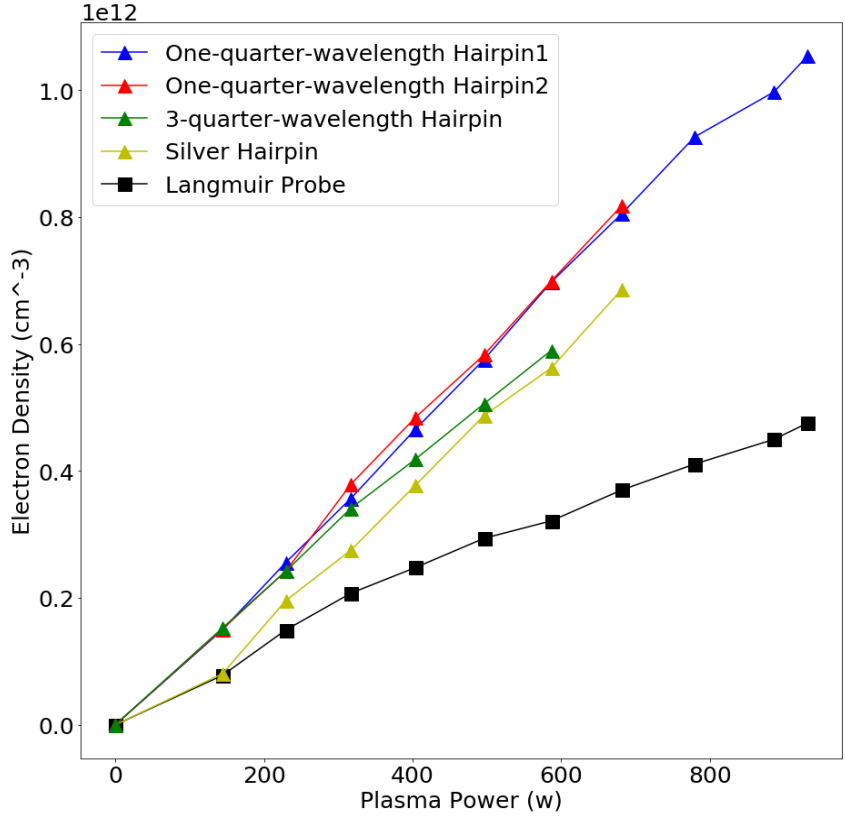

Fig. 13 Sheath corrected density vs. plasma power for the four hairpins described in Table 2. Langmuir probe measurements based on ion saturation current are shown as a comparison.

*4.2 Discussion*

Demonstrated by years of research, the hairpin probe exhibits high accuracy and sensitivity in the low electron density range. We extend the measured electron density to $10^{12}/cm^3$ and the hairpin probe still behaves effectively. In order to measure higher resonant frequencies, the dimensions of the probe must be reduced. Although the quality factor *Q* is not as high as for probes used at lower electron densities, it can be improved by choosing the material with higher conductivity, increasing *l* to *w* aspect ratio, and constructing the 3-quarter-wavelength hairpin. The 3-quarter-wavelength modification doubled the *Q* value, although drawbacks are: First, the signal strength is weaker, reducing the signal to noise ratio. Second, the density measurement range is narrower than one-quarter-wavelength hairpin given the same vacuum resonant frequency. Since the signal loss grows as the frequency increases, it is not a good idea to choose a probe with an unnecessarily high resonant frequency. A desirable choice of the vacuum resonant frequency of the probe is the one slightly higher than the plasma frequency. Coated and uncoated hairpin probes are tested in the experiment. The protection coating has some benefits in hostile plasma environments, although the coated hairpin requires an extra correction. Another advantage of the coating is that it reduces the photoelectron emission from probe surface [10]. In terms of the experimental method, the measurements can be improved by using an IQ mixer. Instead of the frequency-dependent phase being detected, the IQ mixer generates pairs of signals with 90 degrees phase difference. Regarded as sin and cos components, the signal pairs can produce a series of positive signals. The result will be a smooth profile, instead of pulses. Fewer data points will be dropped, and the data analysis process will be easier.

## 5. Conclusion

In this paper, we explore the application of hairpin resonator in measuring electron density in low pressure (20mTorr) inductively coupled plasma (ICP). The measurable density is up to $10^{12}/cm^3$. Various aspects about the design of the system, data analysis and correction and *Q* value of the probe have been discussed. Some improvements are put forward according to observation and calculation. Although the comparison between hairpin probe and Langmuir probe data has been completed, the comparison with the interferometer density measurement is still in process. In the hairpin construction, the using of epoxy may affect the applicability of the simple transmission line model. We believe that this affect will be revealed by the comparison with the interferometer data, and a new model will be introduced in the future.

**Acknowledgement**

Special thanks to staffs at Plasma Science and Technology Institute. And thank Jia Han for helping us set up the apparatus and acquire data.